\documentstyle[12pt]{article}

\def\NPB#1#2#3{{\em Nucl. Phys.} {\bf B#1} (19#2) #3}
\def\PLB#1#2#3{{\em Phys. Lett.} {\bf B#1} (19#2) #3}

\def\PRD#1#2#3{{\em Phys. Rev.} {\bf D#1} (19#2) #3}
\def\PRL#1#2#3{{\em Phys. Rev. Lett.} {\bf#1} (19#2) #3}

\def\ZPC#1#2#3{{\em Zeit. f\"ur Physik} {\bf C#1} (19#2) #3}

\def\IJMPA#1#2#3{{\em Int. J. Mod. Phys.} {\bf A#1} (19#2) #3}
\def\NC#1#2#3{{\em Nuovo Cim.} {\bf A#1} (19#2) #3}

\begin{document}
\begin{titlepage}
\begin{center}
\hfill    IFT--P.027/98\\
~{} \hfill hep-ph/9805329\\

\vskip .1in

{\large \bf Limits on a Strong Electroweak Sector from 
            $e^+ e^- \rightarrow \gamma \gamma + E \!\!\!/ $ at LEP2 }

\vskip 0.3in

Alfonso R. Zerwekh\footnote{zerwekh@axp.ift.unesp.br} and
Rogerio Rosenfeld\footnote{rosenfel@axp.ift.unesp.br}
\vskip 0.1in

{\em Instituto de F\'{\i}sica Te\'orica - Universidade Estadual Paulista\\
     Rua Pamplona, 145 - 01405--900 S\~ao Paulo - SP, Brazil}

\end{center}

\vskip .1in

\begin{abstract}
We study the process $e^+ e^- \rightarrow \gamma \gamma \nu \bar{\nu}$ in the 
context
of a  strong electroweak symmetry breaking model, which can be a source 
of events with two photons and missing energy at LEP2. 
We investigate bounds on the 
model assuming that no deviation is observed from the Standard Model
within a given experimental error.
\end{abstract}
\end{titlepage}

\newpage


Recently there has been a great deal of interest in events with two photons 
plus missing energy at LEP2 because they would be an interesting signature of 
weak scale supersymmetry. 
This signal arises from either $e^+ e^- \rightarrow 
\tilde{N_2} (\tilde{N_2} \rightarrow \tilde{N_1} \gamma) 
\tilde{N_2} (\tilde{N_2} \rightarrow \tilde{N_1} \gamma) $ in models with
gravitational mediated SUSY breaking where the neutralino $\tilde{N_1}$ is 
the lightest
superpartner (LSP)   or
$e^+ e^- \rightarrow 
\tilde{N_1} (\tilde{N_1} \rightarrow \tilde{G} \gamma) 
\tilde{N_1} (\tilde{N_1} \rightarrow \tilde{G} \gamma) $  
in gauge mediated 
supersymmetry breaking where the gravitino $\tilde{G}$ is the 
LSP \cite{susy}. 
Searches for these events have been performed at LEP2 \cite{LEP2} and the 
results are consistent with the Standard Model background once initial state
radiation is taken into account \cite{Mrenna}. 

Supersymmetric models are considered for many reasons the favorite candidates
to extend the extremely successful Standard Model to higher energies.
Among these reasons one could mention: gauge couplings unification, the 
existence of a natural candidate for dark matter of the Universe, a solution
to the naturalness problem and the fact that the theory is weakly coupled
and perturbation theory can be used.
However, there is a logical possibility that the Standard Model is an effective 
theory, being the low energy limit of a {\it strongly coupled} more 
fundamental theory, in the same manner that the non-linear $\sigma$ 
model describes QCD at low energies, in the non-perturbative regime 
where pions are the relevant degrees of freedom \cite{dsb}. Only experiments
will tell us which way Nature chose.

In this letter we show that models of strong electroweak symmetry breaking 
could also be a source of events with two photons and missing energy at
LEP2. We obtain constraints in the
parameter space of one such model from requiring that 
the deviations induced by this model to be smaller than the experimental
accuracy of the measurements . These constraints
are complementary to the ones resulting from precision measurements at LEP.

 
Here we will work in the context of the BESS 
(Breaking Electroweak Symmetry Strongly)\cite{Bess} model.
It can be viewed as an 
effective lagrangian description of the electroweak symmetry breaking due to
an hypothetical strongly interacting sector at the TeV scale. 
The model is based  on the group 
$G'=\left(SU(2)_L \otimes SU(2)_R\right)_{\mbox{global}} 
\otimes \left(SU(2)_V\right)_{\mbox{local}}$. Three new vector bosons
are introduced through the so-called hidden symmetry $SU(2)_V$. 
The group $G'$  breaks down spontaneously to its diagonal subgroup of 
$SU(2)$ giving
rise to six Goldstone bosons. Three of them are absorbed by the new vector 
bosons. 
The remaining three Goldstone bosons give masses to the usual $W^{\pm}$ and 
$Z^{0}$
bosons when the symmetry $SU(2)_L \otimes U(1)_Y \subset 
SU(2)_L \otimes SU(2)_R $ is gauged.

The bosonic part of the lagrangian 
takes the form:
\begin{equation}
{\cal L}=-\frac{v^2}{4}\left[Tr(\tilde{W}-Y)^2 +
\alpha Tr(\tilde{W}+Y-2\tilde{V})^2 \right] +
{\cal L}^{kin}(Y,\tilde{W},\tilde{V})
\end{equation} 
where  $Y,\tilde{W},\tilde{V}$ are the gauge fields associated to 
$U(1)_Y$,  
$SU(2)_L$ and $SU(2)_V$ respectively and $\alpha$ is an arbitrary 
parameter.

A direct coupling to the fermionic sector can be introduced 
through the lagrangian:
\begin{equation}
\begin{array}{ccl}
{\cal L}_{f}&=&\bar{\psi}_{L}i\gamma^{\mu}\left(\partial_{\mu} +
\frac{i}{2(1+b)}g\tilde{W}_{\mu}^{a}\tau^{a} +
\frac{ib}{4(1+b)}g''\tilde{V}_{\mu}^{a}\tau^{a}+
\frac{i}{2}g'y\tilde{Y}_{\mu} \right) \psi_{L}  +\\
            & &\bar{\psi}_{R}i\gamma^{\mu}\left(\partial_{\mu} +
\frac{i}{2}g'y\tilde{Y}_{\mu} \right) \psi_{R}            
\end{array}
\end{equation}
where $b$ is a free parameter. 

These vector bosons are not the physical ones because there are
mixing terms in the lagrangian and the physical gauge bosons are
obtained by diagonalizing the mass matrix in the neutral and charged 
sectors.
After the diagonalization, only the couplings  of the 
physical new vector bosons $V$ to the fermions depend on the $b$ 
parameter.
The physical fields ($A$,$W$,$Z$ and $V$)
are linear combinations of  
$Y,\tilde{W}$ and $\tilde{V}$, and therefore 
the physical $V$ bosons also acquire an indirect coupling to fermions.

The Standard Model is recovered from the BESS model in 
the limit $g''\rightarrow \infty$ and $b \rightarrow 0$,
where $g''$ is the new coupling constant
of $SU(2)_V$.

The model described here is minimal in the sense that only vector resonances 
are introduced. 
Many generalization of this model have been proposed, for example 
models that introduce axial-vector as well as vector particles \cite{Dbess}.
Recently, a model with vector, axial-vector and scalar resonances has
also been studied \cite{Rbess}.

Bounds on this model have been obtained from
precision measurements at LEP1 \cite{Lepbounds}. For a recent review
on results of this model and its extensions see ref. \cite{Frules2}.


The minimal BESS model has three independent free
parameters, which we choose to be $M_V$ (which is given by 
$M^2_V=\alpha\frac{v^2}{4}g''$ in the limit 
$g''\rightarrow \infty$)
$g''$ and  $b$. 
Our calculations show that our results  have little sensitivity in 
$M_V$ as long as $M_V \gg
\sqrt{s}$, so we will use $M_V=400$ GeV in the following.

The Feynman rules for this model are similar to the Standard Model ones with 
modified coupling constant (which can be  found in references \cite{Frules2} and
 \cite{Frules1}).
We included the BESS model particles and couplings into the package
COMPHEP \cite{Comp} and we calculated the cross section for the process 
$e^+ e^- \rightarrow \gamma \gamma \nu \bar{\nu}$ at $\sqrt{s}= 194$ GeV, 
summing over neutrino species,
in the Standard Model and
in the BESS model for different values of $g/g''$ and $b$.

We adopted the following ``loose'' cuts in order to maximize the number of 
detected events \cite{opal}:

\begin{equation}
|\cos(\theta_{\gamma})|<0.7
\end{equation}

\begin{equation}
E_{\gamma}>1.75 \mbox{GeV}
\end{equation}
where $\theta_{\gamma}$  is the angle between the photon 
and the beam, and  $E_{\gamma}$ is the energy of the photon.

We define the quantity

\begin{equation}
\delta\sigma=\frac{\sigma_{BESS}-\sigma_{SM}}{\sigma_{SM}} 
\end{equation}
where $\sigma_{BESS}$ and $\sigma_{SM}$ are the total cross section
predicted by the BESS model and the Standard Model respectively. 
The quantity $\delta\sigma$ measures 
the relative deviation from the Standard Model prediction.
It should be largely insensitive to initial state radiation corrections
and we use $\delta\sigma$ to obtain our results.
 

In order to obtain bounds on the model, we require that
no deviation is observed from the Standard Model prediction
within the experimental error.
Due to runs with small luminosity at different energies, the number 
of events collect so far is very limited and the current 
experimental errors on the cross section measurement ranges 
from $40 \%$ to $100 \%$ \cite{opal}.
Expecting that the measurements will get more accurate for the next run
due to its increased luminosity,
we chose to show our bounds for $\delta\sigma < 0.20, 0.40$
and $0.80$ in figure $1$, where we also include the limits obtained
from precision measurements at LEP \cite{Frules2}.
We can see that the the bounds are complementary and that a 
measurement of the process 
$e^+ e^- \rightarrow \gamma \gamma + E \!\!\!/ $ with larger statistics
can significantly reduce the parameter space available for the BESS
model.

In conclusion, we have shown that models of strong electroweak symmetry 
breaking can also 
be a source of events with two photons plus missing energy at LEP2.
The bounds obtained by requiring no deviation from the Standard Model 
prediction within the experimental error are complementary to the bounds
arising from precision measurements.

\vspace{2cm}
{\Large \bf Acknowledgments}

\vspace{1cm}
We would like to thank Alexander Belyaev for teaching us how to use
Comphep and for useful conversations.
This work was supported by Conselho Nacional de Desenvolvimento 
Cient\'{\i}fico e Tecnol\'ogico (CNPq) and Funda\c{c}\~ao de Amparo \`a 
Pesquisa do Estado de S\~ao Paulo (FAPESP). 
 
\newpage

{\Large \bf Figure Caption}

\vspace{2cm}
{\bf Figure 1:}\\
Limits on the parameter space $g/g'' \times b$ of the BESS model.
The region outside the solid lines is excluded by precision measurements
at LEP. The regions below the dashed, dotted and dot-dashed lines are
excluded  by requiring that
no deviation is observed from the Standard Model prediction for the
process $e^+ e^- \rightarrow \gamma \gamma + E \!\!\!/ $ at LEP2
with $\sqrt{s} = 194$ GeV
within  $20 \%$,$40 \%$ and $80 \%$ accuracy respectively.

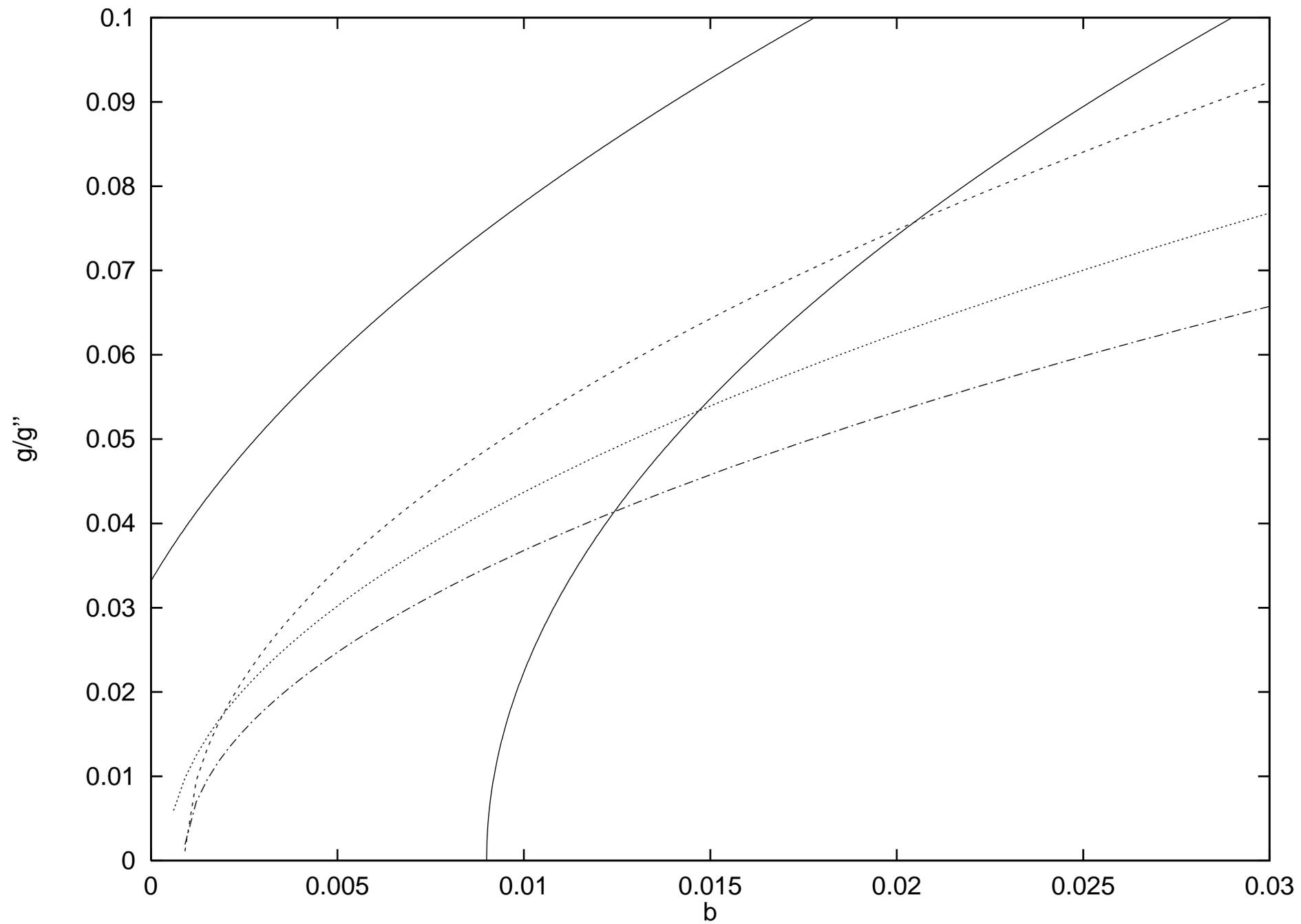
\end{document}